\documentclass[10pt]{article}
\usepackage[cp1251]{inputenc}
\newcommand{\sss}{\scriptstyle}
\usepackage[dvips]{graphicx}   \usepackage[dvips]{epsfig}
\def\lsim{\  \lower-1.2pt\vbox{\hbox{\rlap{$<$}\lower5pt\vbox{\hbox{$\sim$}}}}\ }
\def\gsim{\
  \lower-1.2pt\vbox{\hbox{\rlap{$>$}\lower5pt\vbox{\hbox{$\sim$}}}}\ }

\begin{document}
\title{Quasimomentum of an elementary excitation \\ for a system of point bosons under zero boundary conditions}
\author{ {\small Maksim D. Tomchenko}
\bigskip \\ {\small Bogolyubov Institute for Theoretical Physics  of the NAS
of Ukraine, Kyiv}}
% {\small E-mail:mtomchenko@bitp.kiev.ua}}
% \date{\today}
 \date{\empty}
 \maketitle
 \large
 \sloppy
\textit{Presented by Academician of the NAS of Ukraine V.M. Loktev}
\begin{abstract}
As is known, an elementary excitation of a many-particle system with
boundaries is not characterized by a definite momentum. We obtain
the formula for the \textsl{quasimomentum} of an elementary
excitation for a one-dimensional system of $N$ spinless point bosons
under zero boundary conditions (BCs). In this case, we use the
Gaudin's solutions obtained with the help of the Bethe ansatz. We
have also found the dispersion laws of the particle-like and
hole-like excitations under zero BCs. They coincide with the known
dispersion laws obtained under periodic BCs.
\end{abstract}
% UDC 538.941  \\

The theory of point bosons
\cite{girardeau1960,ll1963,lieb1963,yangs1969,gaudin1971,takahashi1999}
based on the Bethe ansatz is a valuable part of the physics of
many-particle systems, since the system of equations for
quasimomenta $k_{j}$ can be solved exactly at any coupling constant
$\gamma$, and the thermodynamic quantities can be determined from
the Yang--Yang's equations \cite{yangs1969} at any temperature. This
allows one to test the solutions for real nonpoint bosons, the
equations for which can rarely be solved.

In the present work, we will study a one-dimensional (1D) system of
spinless point bosons in the exactly solvable approach,  based on
the Bethe ansatz. For the real systems the boundary conditions (BCs)
are closer to the zero ones ($\Psi(x_{1},\ldots,x_{N})=0$ on the
boundaries), than to the periodic BCs. Therefore, it is of
importance to find the ground-state energy and the dispersion law
under the zero BCs. The ground state was already studied
\cite{gaudin1971,mt2015}, but the dispersion law was not found. To
find it, one needs to determine the energy and the quasimomentum of
a quasiparticle. These problems will be considered in our work. The
main difficulty consists in obtaining the formula for the
quasimomentum, because the ordinary method with the use of the
operator of momentum fails under the zero BCs.

Under the periodic BCs \cite{ll1963}, a quasiparticle possesses the
momentum \cite{lieb1963,takahashi1999,batchelor2006,lang2016}
 \begin{equation}
p=\sum\limits_{i=1}^{N}(\acute{k}_{i}-k_{i}),
     \label{1} \end{equation}
where $k_{i}$ are the solutions for the ground state, and
$\acute{k}_{i}$ are the solutions for the state with one
quasiparticle. This definition of the momentum of a quasiparticle is
self-consistent: the thermodynamic velocity of sound
($v_{s}^{th}=\sqrt{m^{-1}\partial P/\partial \rho}, P=-\partial
E_{0}/ \partial L$, $\rho = N/L$) coincides with the microscopic one
($v_{s}^{mic}=\partial E(p)/\partial p |_{p\rightarrow 0}$)
\cite{lieb1963}.

Under the zero BCs, the quasimomentum of a quasiparticle was
obtained similarly to (\ref{1}) \cite{mt2015,ying2001}:
 \begin{equation}
p=\sum\limits_{i=1}^{N}(|\acute{k}_{i}|-|k_{i}|).
     \label{1-2} \end{equation}
However, in such approach the equality $v_{s}^{th}= v_{s}^{mic}$ is
strongly violated \cite{mt2015}. Below we will define the quantity
$p$ in such a way that this difficulty disappears.

\textbf{Initial equations.} Consider $N$ spinless point bosons
placed on a line of length $L$. The Schr\"{o}dinger equation for
such system reads
\begin{equation}
 -\sum\limits_{j}\frac{\partial^{2}}{\partial x_{j}^2}\Psi + 2c\sum\limits_{i<j}
\delta(x_{i}-x_{j})\Psi=E\Psi.
     \label{2-1} \end{equation}
We use the units with $\hbar=2m=1$. Under the periodic BCs, for each
of the domains $x_{1}\leq x_{2}\leq\ldots \leq x_{N}$ a solution of
the Schr\"{o}dinger equation is the Bethe ansatz
\cite{ll1963,gaudin1971}
\begin{equation}
 \psi_{\{k \}}(x_{1},\ldots,x_{N})=\sum\limits_{P}a(P)e^{i\sum\limits_{l=1}^{N} k_{P_{l}}x_{l}},
      \label{2-2} \end{equation}
where  $k_{P_{l}}$ is one of $k_{1},\ldots,k_{N}$, and $P$ means all
permutations of $k_{l}$. Under the zero BCs, the solution is a
superposition of  counter-waves \cite{gaudin1971}:
\begin{equation}
 \Psi_{\{|k| \}}(x_{1},\ldots,x_{N})=\sum\limits_{\{\varepsilon \}}C(\varepsilon_{1},\ldots,\varepsilon_{N})\psi_{\{k
 \}}(x_{1},\ldots,x_{N}),
      \label{2-20} \end{equation}
where $k_{j}=\varepsilon_{j}|k_{j}|$, $\varepsilon_{j}=\pm 1$. Under
any BCs, the energy of the system is
\begin{eqnarray}
 E=k_{1}^{2}+k_{2}^{2}+\ldots +k_{N}^{2}.
     \label{2-3} \end{eqnarray}
Under the periodic BCs, $k_{j}$ satisfy the Lieb--Liniger's
equations \cite{ll1963} that are usually written in the Yang--Yang's
form \cite{yangs1969}
\begin{eqnarray}
Lk_{i}=2\pi
I_{i}-2\sum\limits_{j=1}^{N}\arctan{\frac{k_{i}-k_{j}}{c}}, \quad
i=1,\ldots, N.                         \label{2-yy} \end{eqnarray}
We will use the Lieb--Lininger's equations in the Gaudin's form
\cite{gaudin1971}:
\begin{eqnarray}
Lk_{i}=\left. 2\pi
n_{i}+2\sum\limits_{j=1}^{N}\arctan{\frac{c}{k_{i}-k_{j}}}\right|_{j\neq
i}, \ i=1,\ldots, N,
   \label{2} \end{eqnarray}
where $n_{i}$ are integers.  For the ground state of the system,
$n_{i}=0$ for all $i=1,\ldots,N$. The systems of equations
(\ref{2-yy}) and (\ref{2}) are equivalent \cite{gaudin1971}. In this
case, $I_{i}=n_{i}+i-\frac{N+1}{2}$.

Under the zero BCs, $k_{j}$ satisfy the Gaudin's equations
\cite{gaudin1971}:
\begin{eqnarray}
L|k_{i}|=\pi n_{i}+\sum\limits_{j=1}^{N}\left
(\arctan{\frac{c}{|k_{i}|-|k_{j}|}} +
\arctan{\frac{c}{|k_{i}|+|k_{j}|}}\right )|_{j\neq i}, \quad
i=1,\ldots, N,
     \label{2-4} \end{eqnarray}
where $n_{i}$ are integers, $n_{i}\geq 1$
\cite{gaudin1971,mtjpa2017}. The ground state corresponds to $n_{i}=
1$ for all $i$. We denote $\rho=N/L$, $\gamma =c/\rho$.

Equations (\ref{2}) has the unique real solution $\{k_{i}\}$
\cite{takahashi1999}, and equations (\ref{2-4}) have the unique real
solution $\{|k_{i}|\}$ \cite{mtjpa2017}.

The quasiparticles are commonly described with the help of the
Yang--Yang's $I_{i}$-numbering (\ref{2-yy}). Below we will introduce
the quasiparticles with the help of the Gaudin's $n_{i}$-numbering
(\ref{2}), (\ref{2-4}), since this way is simpler and more physical
\cite{mtholes}, and allows one to sight the Bose properties of
quasiparticles \cite{mt2015}. These two ways of introduction of
quasiparticles are equivalent. For example, under the periodic BCs,
the ``particle''
$\{I_{i}\}=(1-\frac{N+1}{2},\ldots,N-1-\frac{N+1}{2},N-\frac{N+1}{2}+j)$
with the help of the $n_{i}$-numbering is written as
$\{n_{i}\}=(0,\ldots,0,j)$. In the $n_{i}$-language, the ``hole''
$\{I_{i}\}=(1-\frac{N+1}{2},\ldots,N-2-\frac{N+1}{2},N-\frac{N+1}{2},N+1-\frac{N+1}{2})$
is $\{n_{i}\}=(0,\ldots,0,1,1)$. A way of introduction of
quasiparticles  with the help of the $n_{i}$-numbering was proposed
in \cite{mt2015}.

\textbf{Definition of the quasimomentum of an elementary
excitation.} We now find how the quasimomentum of an elementary
excitation can be determined under the zero BCs. Under the periodic
BCs, the relation \cite{ll1963}
\begin{equation}
\sum\limits_{j=1}^{N}\left (-i \frac{\partial}{\partial x_{j}}\right
)\psi_{\{k \}}(x_{1},\ldots,x_{N})= \left
(\sum\limits_{j=1}^{N}k_{j}\right )\psi_{\{k \}}(x_{1},\ldots,x_{N})
     \label{3-00} \end{equation}
holds in the whole domain $x_{1},\ldots,x_{N}\in [0,L].$ Therefore,
the system has the total momentum
\begin{equation}
P= \sum\limits_{j=1}^{N}k_{j},
     \label{3-0} \end{equation}
and the momentum of a quasiparticle is given by formula (\ref{1}).
Under the zero BCs, the relation
\begin{eqnarray}
\sum\limits_{j=1}^{N}\left (-i \frac{\partial}{\partial x_{j}}\right
)\Psi_{\{|k| \}}(x_{1},\ldots,x_{N})=
f(|k_{1}|,\ldots,|k_{N}|)\Psi_{\{|k| \}}(x_{1},\ldots,x_{N})
\nonumber \end{eqnarray} is not satisfied. Therefore, the system has
no definite momentum. To find the formula for the quasimomentum of
an excitation, we use the following property. It is known that the
momentum (quasimomentum) of a quasiparticle is quantized by the law
$p_{j}=\hbar 2\pi j/L$ ($j=\pm 1, \pm 2, \ldots$) under the periodic
BCs \cite{bog1947}  and $p_{j}=\hbar \pi j/L$ ($j=1,  2, \ldots$)
under the zero BCs \cite{cazalilla2004,mtujp2019}. Starting from
these relations, one can guess the formula for the momentum
(quasimomentum).

Consider a periodic system. Equations (\ref{2}) yield
\begin{eqnarray}
\sum\limits_{j=1}^{N}k_{j}=
\frac{2\pi}{L}\sum\limits_{j=1}^{N}n_{j}.
     \label{3-1} \end{eqnarray}
It is seen that the quantity $P=\sum_{j=1}^{N}k_{j}$ is quantized in
the same way as the momentum of an ensemble of quasiparticles
\cite{bog1947}. Therefore, it is natural to identify $P$ with the
total momentum of the system (in the reference system, where the
center of masses is at rest). We obtain that
$P_{0}=\sum_{j=1}^{N}k_{j}=0$ is for the ground state and
$P_{1}=\sum_{j=1}^{N}k_{j}=2\pi r/L$ for the state with one
particle-like excitation ($n_{i\leq N-1}=0$, $n_{N}=r\neq 0$). The
momentum of a particle-like excitation
 \begin{equation}
p=P_{1}-P_{0}=\sum\limits_{i=1}^{N}(\acute{k}_{i}-k_{i})=\frac{2\pi
n_{N}}{L}=\frac{2\pi r}{L}
     \label{3-2} \end{equation}
corresponds to formula (\ref{1}) and to momentum quantization
$p_{j}= 2\pi j/L$ \cite{bog1947}. We have solved system (\ref{2})
numerically, found the energies of the ground and excited states,
and obtained that the equality $v_{s}^{th}= v_{s}^{mic}$ holds with
high accuracy: for $\rho=1, N=200, 1000, 5000$ and $\gamma=0.1, 1,
10,$ the equality $v_{s}^{th}= v_{s}^{mic}$ holds with an error of
$\lsim 0.1\%$. In this case, the error depends strongly on $\gamma$
and $N$: $\frac{|v_{s}^{mic}-v_{s}^{th}|}{v_{s}^{th}}\simeq
\frac{0.01}{\gamma N}$.

We now consider  the system under the zero BCs. Relation (\ref{2-4})
yields
\begin{eqnarray}
\sum\limits_{j=1}^{N}|k_{j}|=
\frac{\pi}{L}\sum\limits_{j=1}^{N}n_{j}+\frac{1}{L}\sum\limits_{i,j=1}^{N}
\arctan{\frac{c}{|k_{i}|+|k_{j}|}}|_{j\neq i}.
     \label{3-3} \end{eqnarray}
Introduce the quantity
\begin{eqnarray}
P(\{|k_{i}|\})=\sum\limits_{j=1}^{N}|k_{j}|-\frac{1}{L}\sum\limits_{l,j=1}^{N}
\arctan{\frac{c}{|k_{l}|+|k_{j}|}}|_{j\neq l},
     \label{3-4} \end{eqnarray}
then relations (\ref{3-3}) and (\ref{3-4}) yield
\begin{eqnarray}
P(\{|k_{i}|\})=\frac{\pi}{L}\sum\limits_{j=1}^{N}n_{j}.
     \label{3-4b} \end{eqnarray}
Since $P$ (\ref{3-4}), (\ref{3-4b}) is quantized similarly to the
quasimomentum of the ensemble of quasiparticles for an interacting
system under the zero BCs \cite{mtujp2019}, it is natural to
identify $P$ (\ref{3-4}), (\ref{3-4b}) with this quasimomentum. It
is essential that the quasiparticles  are introduced for a system of
point bosons in such a way that the total number of quasiparticles
is $\leq N$ (the same limitation exists also for a system of
nonpoint bosons \cite{mtholes}). This limitation agrees with
(\ref{3-4b}). The smallest quasimomentum of the system corresponds
to the ground state:
\begin{equation}
P_{0}=P(n_{i\leq N}=1)=\frac{\pi}{L}\sum\limits_{j=1}^{N}1=\frac{\pi
N}{L}=\pi \rho.
 \label{3-p0} \end{equation}
The quasimomentum of a particle-like excitation is
\begin{eqnarray}
&& p_{r-1}=P(n_{i\leq N-1}=1,n_{N}=r)-P(n_{i\leq N}=1)=\nonumber
\\ && =\sum\limits_{j=1}^{N}(|\acute{k}_{j}|-|k_{j}|)-\frac{1}{L}\sum\limits_{l,j=1}^{N}
\left (\arctan{\frac{c}{|\acute{k}_{l}|+|\acute{k}_{j}|}}-
\arctan{\frac{c}{|k_{l}|+|k_{j}|}}\right )|_{j\neq l},
     \label{3-5} \end{eqnarray}
where $\{|\acute{k}_{j}|\}$ and $\{|k_{j}|\}$ are solutions of
Gaudin's equations (\ref{2-4})  for the states with one
particle-like excitation and without excitations, respectively.
Relations (\ref{3-4b}), (\ref{3-5}) yield
\begin{equation}
p_{r-1}=\pi (r-1)/L,
     \label{3-6} \end{equation}
where $r$ is equal to the value of $n_{N}$ for the state with one
particle-like excitation: $r=n_{N}=2,3,4,\ldots; n_{j\leq N-1}=1$.
We have obtained the quantity with the required law of quantization:
$p_{j}=\pi j/L$ \cite{cazalilla2004,mtujp2019}. The numerical
analysis has shown that the equality $v_{s}^{th}= v_{s}^{mic}$ is
satisfied with an error of $\lsim 1\%$ for $\rho=1$; $\gamma=0.1, 1,
10$; $N=200, 1000, 5000$. This error depends on $\gamma$ and $N$
approximately as $\frac{|v_{s}^{mic}-v_{s}^{th}|}{v_{s}^{th}}\simeq
\frac{0.5}{\sqrt{\gamma} N}$. In this case, the linearity of the
dispersion law requires $\sqrt{\gamma} N\gg 1$.  It is significant
that, for the zero and periodic BCs, the error  disappears as
$N\rightarrow \infty$.  That is, this error is due to the finiteness
of a system (for very large $N$ one more  error, related to a
numerical method, should  appear). The equality $v_{s}^{th}=
v_{s}^{mic}$ must be exact in the thermodynamic limit and may be
violated for not large $N, L$. Thus, in the thermodynamic limit, our
formulae agree with the exact equality $v_{s}^{th}= v_{s}^{mic}$.
Hence, formulae (\ref{3-5}) and (\ref{3-6}) for the quasimomentum
are exact, at least as $N, L\rightarrow \infty$.

We note that, for the zero BCs, the error is larger by $1$--$2$
orders of magnitude, than in the periodic BCs case. We suppose that
this is connected with a nonuniformity of the wave function near
boundaries. In particular, for a periodic system, the solution for
the ground-state energy $E_{0}$ becomes close to Bogoliubov's
asymptotic solution $E_{0}(N\rightarrow \infty)$ \cite{bog1947}, if
$N\gsim 100$; for the zero BCs, this occurs for larger $N$: $N\gsim
1000$.

Thus, we have obtained the formula for the quasimomentum of a
quasiparticle  for the system under the zero BCs. Apparently,
quasimomentum (\ref{3-4}), (\ref{3-4b}) corresponds to an accidental
integral of motion. It would be of interest to clarify which
operator corresponds to the quasimomentum (\ref{3-4}).

Let us find the dispersion law $E(p)$ of particle-like excitations
for a system under the zero and periodic BCs. Under the zero BCs we
are based on (\ref{3-6}) and the formula for the energy of a
quasiparticle is \cite{lieb1963}
\begin{equation}
E=\sum\limits_{j=1}^{N}(\acute{k}^{2}_{j}-k^{2}_{j}).
     \label{3-7} \end{equation}
Under the periodic BCs we use formulae (\ref{3-2}), (\ref{3-7}). We
find the solutions $\{\acute{k}_{j}\}$ and $\{k_{j}\}$ from Eqs.
(\ref{2}) under the periodic BCs and from Eqs. (\ref{2-4}) under the
zero BCs. In this case, $\{\acute{k}_{j}\}$ corresponds to the state
with one quasiparticle ($n_{j\leq N-1}=0$, $n_{N}=r$ for the
periodic BCs and $n_{j\leq N-1}=1$, $n_{N}=r>1$ for the zero BCs),
whereas $\{k_{j}\}$ corresponds to the ground state ($n_{j\leq N}=0$
for the periodic BCs and $n_{j\leq N}=1$ for the zero BCs). We have
solved Eqs. (\ref{2}), (\ref{2-4}) numerically and determined the
dispersion law $E(p)$ for the zero and periodic BCs. As is seen from
Fig. 1, the dispersion laws $E(p)$ under the periodic and zero BCs
\textit{coincide}. The numerical solution of systems (\ref{2}) and
(\ref{2-4}) indicates that the ground-state energy ($E_{0}$) under
the zero BCs exceeds $E_{0}$ under the periodic BCs by only a small
surface contribution ${\sss \triangle} E_{0}\sim E_{0}/N$
\cite{mt2015}. For interacting nonpoint bosons, the picture is
similar: at any repulsive interatomic potential, the values of
$E_{0}$ and $E(p)$ of a 1D system under the zero BCs
\cite{mtujp2019} coincide with $E_{0}$ and $E(p)$ of the periodic
system \cite{bog1947}. Moreover, for a 1D system of interacting
bosons it was found in the harmonic-fluid approximation that the
sound velocity is identical under the periodic and zero BCs
\cite{cazalilla2004}.

\begin{figure}[ht]
 % \vspace*{7.5cm}
\centerline{\includegraphics[width=85mm]{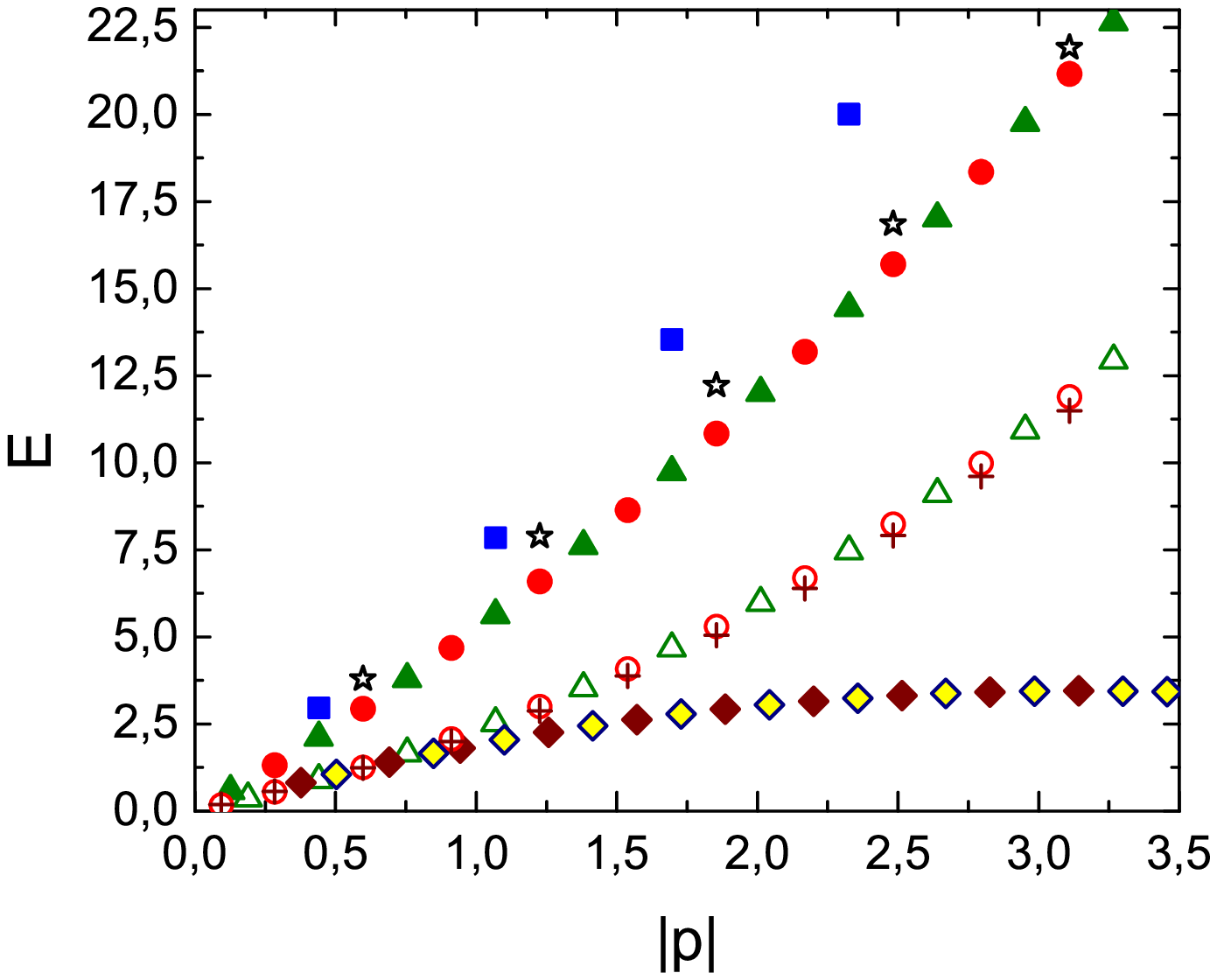} } \caption{
[Color online] Dispersion curves $E(p)$ obtained by the numerical
solution of Eqs. (\ref{2}), (\ref{2-4}) within the Newton method for
$N=L=1000$. 1) $\gamma=1$: $E(p)$ of particle-like excitations under
the periodic BCs (open circles), under the zero BCs (open
triangles), and the Bogoliubov law \cite{bog1947}
$E=\sqrt{p^4+4\gamma \rho^{2}p^{2}}$ (crosses); 2) $\gamma=10$:
$E(p)$ of particle-like excitations under the periodic BCs
(circles), under the zero BCs (triangles), the Bogoliubov law
(stars), and the Girardeau's law \cite{girardeau1960} $E=p^{2}+2\pi
\rho|p|$ (squares);  3) $\gamma=1.725$: $E(p)$ of hole-like
excitations under the periodic (open diamonds) and  zero (diamonds)
BCs.
 \label{fig1}}
\end{figure}

We have also calculated  the dispersion law of hole-like
excitations. It is seen from Fig.~1 that the dispersion law is the
same under the zero and periodic BCs. Visually, it coincides with
the dispersion law of holes obtained by Lieb \cite{lieb1963}. Under
the zero BCs, holes correspond to the states with the following
quantum numbers $n_{j}$: $n_{1\leq j\leq l}=1, n_{l<j\leq N}=2$,
where $l=0,1,\ldots,N-2$. Under the periodic BCs, holes are the
states with $n_{1\leq j\leq l}=0, n_{l<j\leq N}=1$
($l=0,1,\ldots,N-2$) and the states with $n_{1\leq j\leq k}=-1,
n_{k<j\leq N}=0$ ($k=2,3,\ldots,N$). Formula (\ref{3-4b}) implies
that the quasimomentum of a hole under the zero BCs is $p=\pi
(N-l)/L$; the largest quasimomentum is $p=\pi N/L=\pi\rho$. Under
the periodic BCs, the hole has momentum (\ref{1}), (\ref{3-1}),
which takes values from $p=-2\pi\rho$ to $p=2\pi\rho$. Note that, as
shown in work \cite{mtholes}, a hole is a set of interacting
particle-like excitations.

We note that the formulae for the quasimomentum and the solutions
for the dispersion laws, obtained above under the zero BCs, are new
results.

Interestingly,  the dispersion law of particle-like excitations
(Fig. 1) differs at $\gamma=1$ from the Bogoliubov law only by
$5\%$. In this case, the available criterion of applicability of the
Bogoliubov model in the 1D case  for the zero and periodic BCs is as
follows (at $T=0$) \cite{mtujp2019}:
\begin{equation}
\frac{\sqrt{\gamma}}{2\pi}\ln{\frac{N\sqrt{\gamma}}{\pi}}\ll 1.
     \label{3-8} \end{equation}
According to (\ref{3-8}),  it should be $\gamma\rightarrow 0$ as
$N\rightarrow \infty$. But the solutions  $E_{0}$ and $E(p)$ for
point bosons are close to the Bogoliubov solutions even at
$N\rightarrow \infty$, $\gamma \sim 1$ (as for the periodic BCs, see
\cite{ll1963,lieb1963}; for the zero BCs, it was found \cite{mt2015}
that the solutions  $E_{0}$ and $E(p)$ obtained in the limit
$N\rightarrow \infty$ coincide (with an error of $1\%$)  with
$E_{0}$ and $E(p)$ found by directly numerically solving  Eqs.
 (\ref{2-4}) at $N=1000$; therefore, the
dispersion law $E(p)|_{N\rightarrow \infty}$ coincides with the
above-found one $E(p)|_{N= 1000}$ and is close to the Bogoliubov
law, if $\gamma \lsim 1$). We remark that the dispersion law for
$\gamma=10$ (see Fig. 1) is closer to the Bogoliubov law, than to
the Girardeau's one. Though it would be expected the contrary, since
the Girardeau's formula is exact at $\gamma=+\infty$, whereas the
Bogoliubov formula loses its meaning at such $\gamma$. The reason
for the applicability of the Bogoliubov solutions at not small
$\gamma$ is yet unclear.

It was obtained \cite{mt2015} that the dispersion laws of
particle-like excitations  under the zero and periodic BCs are
strongly different. However, this difference is unphysical: it arose
because, under the zero BCs, formula (\ref{1-2}) was used  instead
of formula (\ref{3-5}).

The question is,  how to measure the dispersion law in a system
under the zero BCs? Apparently, this can be made with the help of an
ordinary scattering. But we do not know how to pass from the
Gaudin's wave function (\ref{2-20}) to a localized wave package with
a definite momentum.

The present work was partially supported by the Program of
Fundamental Research of the Department of Physics and Astronomy of
the National Academy of Sciences of Ukraine (project No.
0117U000240).
% выше - это целевая

     \renewcommand\refname{}

\end{document}